\documentclass[twocolumn,showpacs,preprintnumbers]{revtex4}
\usepackage{graphicx}
\usepackage{dcolumn}
\usepackage{bm}

\input{tcilatex}

\begin{document}

\title{Low temperature superlattice in monoclinic PZT}
\author{B. Noheda, L. Wu, and Y. Zhu}
\affiliation{Brookhaven National Laboratory, Upton, 11973-New York}
\date{\today}

\begin{abstract}
TEM has shown that the strongly piezoelectric material PbZr$_{0.52}$Ti$%
_{0.48}$O$_{3}$ separates into two phases at low temperatures. The majority
phase is the monoclinic phase previously found by x-ray diffraction. The
minority phase, with a nanoscale coherence length, is a slightly distorted
variant of the first resulting from the anti-phase rotation of the oxygen
octahedra about [111]. This work clears up a recent controversy about the
origin of superlattice peaks in these materials, and supports recent
theoretical results predicting the coexistence of ferroelectric and
rotational instabilities.
\end{abstract}

\maketitle

\affiliation{Brookhaven National Laboratory, Upton, 11973- New York}

\affiliation{Brookhaven National Laboratory, Upton, New York 11973,USA.}



Ferroelectric ceramics of PbZr$_{1-x}$Ti$_{x}$O$_{3}$ (PZT) with
compositions around x=0.50 display anomalously high dielectric and
piezoelectric responses, which are related to the "morphotropic phase
boundary" (MPB), the steep boundary separating the rhombohedral (zirconium
rich) and tetragonal (titanium rich) phases of the phase diagram \cite{Jaf1}%
. The technological relevance of PZT as the active element in
electromechanical transducers has motivated a large amount of fundamental
research in the last fifty years, aimed at revealing the nature of the MPB
and the origin of the outstanding physical properties of these materials,
which to this date are still not well understood.

Recently, a monoclinic phase with space group (s.g) Cm, has been discovered
by x-ray powder diffraction at the MPB of PZT \cite{Noh1,Noh2,Noh3}, in
between the rhombohedral (R) and tetragonal (T) phases, as shown in Fig. 1.
The importance of this new phase (called M$_{A}$ after ref.\cite{Van1}) is
remarkable because, due to the lack of a symmetry axis, it allows for the
rotation of the ferroelectric polarization between the polar axes of the R
and T phases \cite{Noh2,Bel1}. The remaining symmetry element is a mirror
plane, the pseudo-cubic ($\overline{1}$10) plane,which is also common to T
and R (with s.g. P4mm and R3m, respectively). Due to the near degeneracy of
the different phases at the MPB, the polarization rotation can also be
easily achieved by applying an electric field, which induces the monoclinic
phase, and thus explains the high electromechanical response observed in PZT 
\cite{Guo1,Bel2}.

First principles calculations have been able to reproduce the intermediate
monoclinic phase observed in PZT in excellent agreement with the
experiments, provided that the atomic disorder in Zr/Ti site is taken into
account \cite{Bel1}. Furthermore, they have shown that this phase is
directly related to the very high electromechanical response of the ceramic
material (single crystals of PZT are not available) mainly due to the d$%
_{15} $ component of the piezoelectric tensor, which indicates the easy
rotation of the polarization in the monoclinic plane \cite{Bel1}. From a
phenomenological point of view, it has recently been shown that the
monoclinic phase can be derived from the Devonshire expansion of the free
energy to eighth-order, while a twelfth-order expansion would be needed to
derive the lowest symmetry triclinic perovskites \cite{Van1}. All the above
is a clear indication of the very high anharmonicity of the energy
potentials in PZT, that is also present in other related systems \cite{Kia1}.

With decreasing temperatures PbZr$_{0.52}$Ti$_{0.48}$O$_{3}$ (PZT48)
transforms from a cubic to a tetragonal phase at about 660K, and from a
tetragonal to a monoclinic phase at about 300K \cite{Noh2,Noh3}. X-ray
diffraction reveals no further phase transformation down to 20K \cite{Noh2}.
The reported M$_{A}$ cell is rotated $45^{o}$ about the $c$-axis with
respect to the tetragonal one and is double in volume, with $a_{m}$ $\simeq $
$b_{m}$ $\simeq $ $a_{p}$$\sqrt{2}$, and $c_{m}$$\simeq $ $a_{p}$, $a_{p}$ $%
\simeq $ 4 \AA \thinspace\ being the length of the cubic cell. However,
recently, Ragini et al. \cite{Rag1} have observed superlattice (sl)
reflections at low temperatures by transmission electron microscopy (TEM)
that are not consistent with the M$_{A}$ phase. These sl reflections are
also observed by neutron diffraction \cite{Noh3,Raj1,Fra1}, but are not seen
in the x-ray diffraction patterns \cite{Noh2,Rag1}.

The appearance of a superlattice is a common phenomenon in pervoskites,
related in most cases to the softening of one or more $\Gamma $$_{25}$
zone-corner (R-point) phonons \cite{Shi1}, which involves rotations of the
oxygen octahedra\cite{Gla2}. In ferroelectric perovskites, the octahedra
tilts occur independently of the cation displacements (associated with the
softening of the $\Gamma $$_{15}$ zone-center mode) and therefore do not
essentially affect the ferroelectric properties. In PZT, such rotations have
been observed in the rhombohedral region of the phase diagram (see Fig. 1) 
\cite{Gla1}. At low temperatures, in the R$_{LT}$ phase, in addition to the
cation displacements along the pseudo-cubic [111] direction, there is a tilt
of the oxygen octahedra about the [111] axis. This rotation doubles the unit
cell \cite{Gla1,Gla2} and produces sl reflections of the $1/2\left\{ {hkl}%
\right\} $ type ($h$,$k$,$l$ all odd) \cite{Note1}, with an intensity
approximately proportional to the tilt angle.

To explain the sl reflections recently observed in monoclinic PZT, Rajan et
al. \cite{Raj1}, based on a Rietveld analysis of neutron powder diffraction
data, have proposed that the M$_A$ phase transforms, at low temperatures,
into a different monoclinic phase, with space group Pc, in which the M$_{A}$
unit cell is doubled along the $c$-direction due to an anti-phase octahedral
tilt about the $c$-axis. However, Rietveld analysis of disordered systems
with low symmetry is not unambiguous due to the number of constraints that
need to be included. For example, similar neutron patterns have also been
successfully modelled by Frantii et al. \cite{Fra1} in terms of the
coexistence of monoclinic M$_{A}$ and rhombohedral R$_{LT}$ phases. In this
letter we clarify this controversy by means of TEM measurements on a PZT48
sample at low temperatures. We show that the M$_{A}$ phase persists at low
temperatures and that the observed superlattice originates from nanoregions
of the sample that undergo rotations of the oxygen octahedra along the [111]
direction, without altering the cation distortion.

TEM experiments were carried out using a JEOL 300kV field-emission
microscope equipped with an energy filter and low temperature stages.
Diffraction and image data were recorded using parallel beam on either
imaging plates or CCD cameras. The same ceramic pellets of ref. \cite{Noh2}
were used in these experiments. The excellent quality of the samples was
established in previous x-ray diffraction work \cite{Noh2} that showed very
narrow Braggs peaks and very sharp and well-defined phase transitions. TEM
samples were prepared using a standard thinning procedure, i.e. first
mechanical polishing down to less than $10\mu m$, then ion-milling to
perforation with low-energy ion-guns. The thickness of the samples usually
ranged from 50-100nm. To minimize multiple scattering, thinner regions
(~5-10nm) were also used. Electron diffraction presents the advantage of
simultaneously acquiring dozens of reflections from a local area and
reaching far out in reciprocal space due to the high energy of the incident
electrons. In the presence of domain variants and twins, formed due to the
reduction of crystal symmetry, electron diffraction can unambiguously reveal
the change of the crystal symmetry, including those caused even by an
extremely small lattice distortion, by the splitting of high-order Bragg
reflections.

Diffraction patterns containing sl reflections of the $1/2\left\{ {hkl}%
\right\} $ ($h,k,l$ all odd) type (pseudocubic indexing will be used unless
stated otherwise) were observed in PZT48 at low temperatures, in agreement
with Ragini et al. \cite{Rag1}. However, such reflections were also found to
vary in intensity along the sample. Figs. 2a-b show two diffraction
patterns, both taken in the pseudo-cubic \TEXTsymbol{<}110\TEXTsymbol{>}
zone of the reciprocal space at 87 K, corresponding to two different sample
areas. While the sl spots are clearly visible in Fig. 2a, they are very weak
and difficult to detect in Fig. 2b. The dark-field images formed by the sl
reflections in both areas are presented in Figs. 2c-d. The dark background
and the bright spots correspond to the simple-lattice and the superlattice,
respectively. These figures clearly show that the sample consist of two
phases and that only one of them displays superlattices, contrary to the
previously proposed models \cite{Raj1}. A series of dark field images shows
that the volume fraction of the superlattice phase varies in different areas
of the sample from 0 to about 30\% of the total volume, clearly showing its
minority character. Furthermore, these images reveal a coherence length as
short as 3nm for the superlattice phase (see Fig. 2c).

Fig. 3a shows the pattern expected in the \TEXTsymbol{<}110\TEXTsymbol{>}
zone of a pseudo-cubic perovskite phase. The tetragonal, T, rhombohedral R$%
_{HT}$ and monoclinic M$_{A}$ phases of PZT show similar patterns, since the
distortion from the cubic phase is very small. None of these phases have a
sl of the type observed in Fig. 2. However, as mentioned above, the R$_{LT}$
phase of PZT (with s.g. R3c) is known to display similar sl reflections \cite
{Vie1,Ric1}, which, together with its proximity in the PZT phase diagram
(see Fig. 1) makes this phase a good candidate to check \cite{Note1}.

Fig.4a shows the diffraction pattern in the <211> zone at around 87 K. It is
seen that both the fundamental and sl reflections split along the <111>
direction (see insets). Dynamic diffraction analysis shows that none of the
twin variants for crystals with a rhombohedral symmetry (s.g. R3c) yield
this type of splitting \cite{Wu1,Ten1} and, thus, it is possible to reject
the presence of the R$_{LT}$ phase. The split is, however, consistent with
the monoclinic distortion. Moreover, the fact that both main and sl
reflections show the same kind of splitting clearly indicates that both
phases share the same fundamental lattice (otherwise extra spots arising
from a second cell would be observed in inset 2).

Fig. 4b shows an electron diffraction pattern in the \TEXTsymbol{<}110%
\TEXTsymbol{>} zone at the same temperature. The main reflections are seen
to split into three spots (see inset 4) consistently with the monoclinic
symmetry, and with the reported M$_A$ phase \cite{Noh2}. It can also be
noticed that the sl spots in this zone do not split (see inset 3), which
indicates that only one of the three observed twins is responsible for the
sl phase and, therefore, confirms the two-phase scenario. Further
information can yet be extracted from the diffraction experiments: The
extintion rules show that the mirror plane of the M$_{A}$ phase is not
present in the sl phase. Furthermore, the experiments give extra information
about the symmetry of this phase by showing that the three-dimensional
reciprocal lattice is face-centered, the real lattice being therefore body
centered.

Although other effects could also produce a superlattice (i.e. cation
ordering or anti-parallel cation displacements.), the fact that the sl
reflections are observed with neutrons and not with x-rays indicates that
they are due to rotations of the oxygen octahedra, for which x-rays are not
very sensitive. Bearing all the above in mind, we propose a model for the
minority sl phase in which the cations keep the M$_{A}$ distortion, while
the oxygen octahedra are rotated in an anti-phase fashion about [111], as in
the R$_{LT}$ phase. Fig. 3e shows the projection of the octahedra framework
on the pseudo-cubic ($\overline{1}$10) mirror plane. The M$_{A}$ lattice
vectors $\overrightarrow{a}_{m}$ and $\overrightarrow{c}_{m}$ are contained
in the plane, while $\overrightarrow{b}_{m}$ is perpendicular to it. After
the tilting, the unit cell doubles along $c$ (dashed lines), similar to the R%
$_{LT}$-R$_{HT}$ phase transition for smaller Ti contents, and the cell
becomes body centered (s.g. Ic). A unit cell can be chosen to keep the
standard c-centered space group Cc as represented by the thick solid lines
in the figure.

The sl reflections expected in the \TEXTsymbol{<}110\TEXTsymbol{>} zone for
the new Cc phase are depicted in Fig. 3b, and are in perfect agreement with
experiment (see Fig. 2) \cite{Note2}. The minority character of the tilted
phase, and the broadening effects associated with the small size of the
tilted regions, explains the low intensity on the sl reflections in the
neutron diffraction patterns \cite{Raj1,Fra1}. Moreover, according to our
model, the structure factor of the $1/2\left\{hhh\right\} $ reflections is
much smaller than that of the $1/2\left\{hkl\right\}$ ones, which also
explains why the $1/2\left\{111\right\}$ sl peak is not observed with
neutrons. The monoclinic space group Pc recently proposed \cite{Raj1}, in
which the oxygen rotations are along the [001] axis, can be discarded since
it would give rise to patterns like those shown in Figs. 3c-d, which contain
sl reflections that were not observed (open circles).

The evolution of the sl reflections was monitored as a function of
temperature. They started to disappear at T= 150 K in certain areas of the
sample, but were still visible in other areas at T= 200K. At about T= 230K
no superlattice could be found, in agreement with Ragini et al. \cite{Rag1}.
This behavior suggests that local internal inhomogeneities or local stresses
(most likely originating from cation disorder) cause phase separation by
favoring the octahedra tilts in certain regions of the sample, at low
temperatures, and that the transition temperature between the tilted and
non-tilted phases depends to a large extent on the local environment. This
is in perfect agreement with calculations by Fornari and Singh, who predict
an instability of the rotational degrees of freedom, comparable to the
ferroelectric one, as well as a strong pressure dependence of these \cite
{For1}. Further studies need to be done to clarify whether the tilts are
associated with Zr/Ti-rich regions.

In summary, our results clearly show that the M$_A$ phase, which is known to
be directly related to the unusual piezoelectric and ferroelectric
properties of PZT and related systems, remains stable at low temperatures,
contrary to recent reports. Some areas of the sample, as small as 3nm,
undergo rotations of the oxygen octahedra about the [111] direction, similar
to those of the neighboring R$_{LT}$ phase, that lower the symmetry but do
not modify the fundamental lattice, therefore keeping the ferroelectric
properties basically unaltered. The temperature evolution of the tilted
regions supports the theoretical results of Fornari and Singh that predict
the possible coexistence of ferroelectric and rotational instabilities due
to local stress fields \cite{For1}.

Acknowledgments

The authors are specially grateful to D.E. Cox, G. Shirane and T. Vogt for
very insightful discussions that have helped improve this manuscript
considerably, to P. Gehring for useful discussions, to J. Frantii for
providing the results of his unpublished neutron diffraction work and for
helpful discussions, to R. Guo and S.-E. Park for supplying the samples, and
to J. Li for assistance in the experiments. Financial support from the U.S.
Department of Energy under contract No. DE-AC02-98CH10886 is also
acknowledged.

\newpage 
\begin{figure}[tbp]
\includegraphics[width=0.9\textwidth] {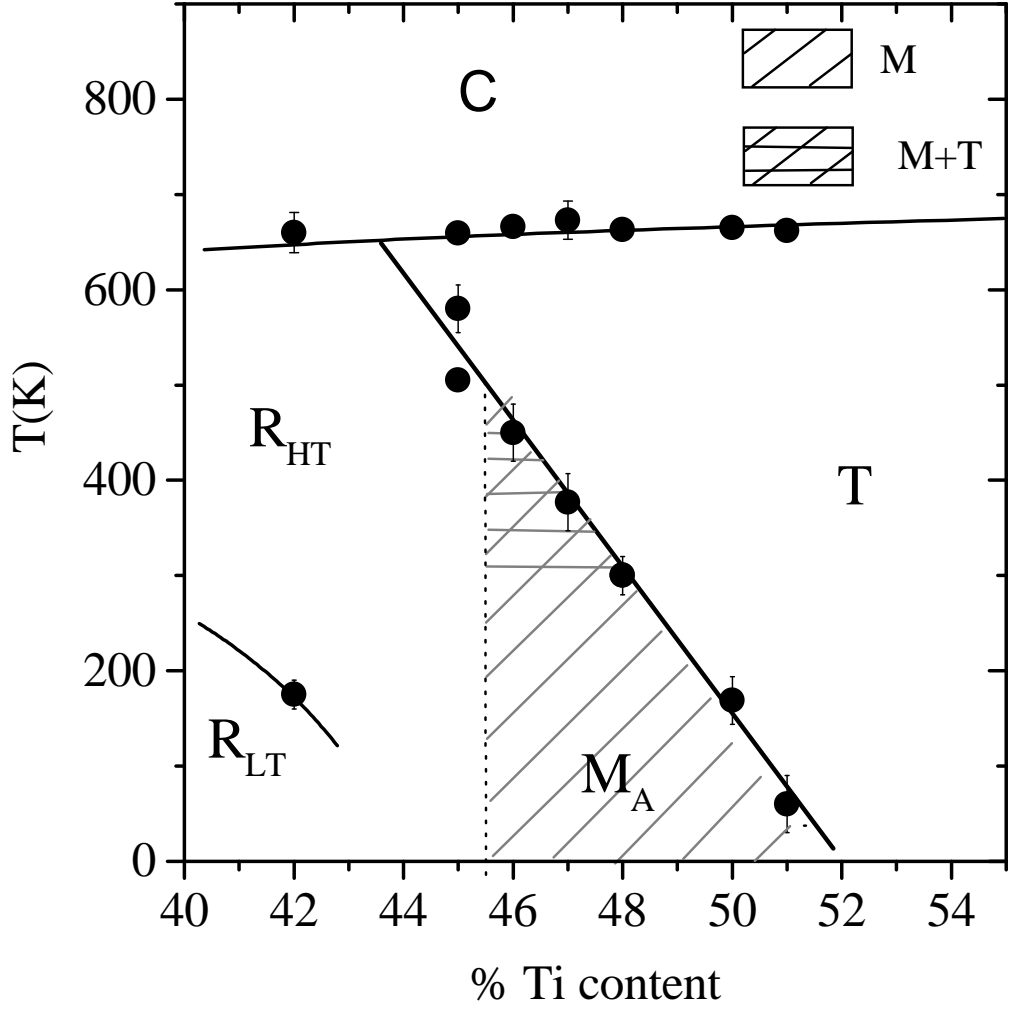}
\caption{Phase diagram of PbZr$_{1-x}$Ti$_{x}$O$_{3}$ (PZT) around the
morphotropic phase boundary adapted from ref.\protect\cite{Noh3}}
\end{figure}

\begin{figure}[tbp]
\includegraphics[width=0.7\textwidth] {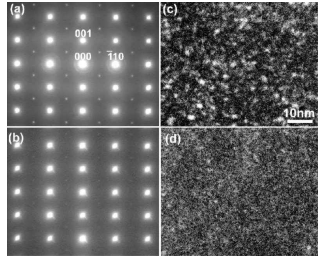}
\caption{Electron diffraction patterns observed in the pseudo-cubic 
\TEXTsymbol{<}110\TEXTsymbol{>} zone of PbZr$_{0.52}$Ti$_{0.48}$O$_{3}$ at
87 K in two different areas, showing strong (a) and weak (b) superlattice
reflections. The dark field images formed by the superlattice reflections
are shown in c) and d) for two regions with strong and weak superlattice
peaks, respectively}
\end{figure}

\begin{figure}[tbp]
\includegraphics[width=0.7\textwidth] {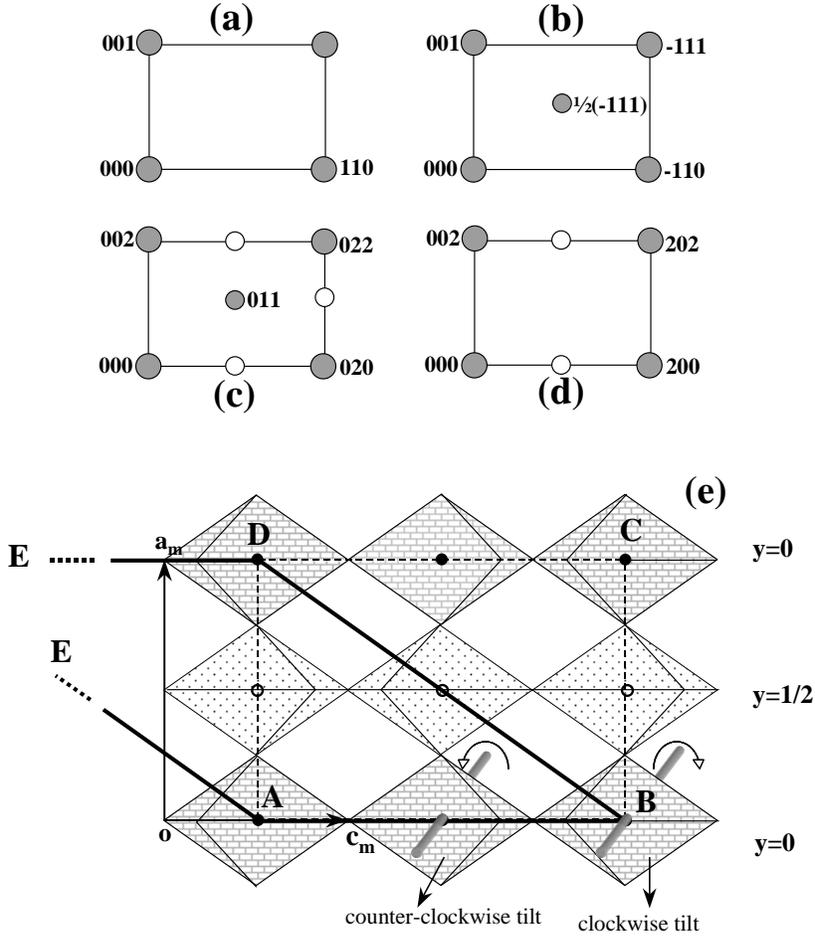}
\caption{Sketch of the reciprocal lattice expected in the pseuco-cubic 
\TEXTsymbol{<}110\TEXTsymbol{>} zone for a) the monoclinic Cm phase; b) the
monoclinic Cc phase; c) and d) the monoclinic Pc phase (here using Pc
indices for clarity). e) Real-space projection of the octahedral framework
on the pseudo-cubic ($\overline{1}$10) plane, showing the tilt pattern
responsible for the observed sl reflections. Oxygen atoms are located at the
vertices of the octahedra, Zr/Ti are located at the center of the octahedra.
Solid and open circles represent Zr/Ti at y=0 and y=1/2, respectively. Pb
atoms are omitted for clarity. The projection of the new Ic unit cell (ABCD)
with a doubled $c$-constant, formed after the anti-phase tilting along the
pseudo-cubic [111] direction, is marked by dashed lines. The Cc unit cell
(ABDE) with $a_{m}^{\prime }$= $\protect\sqrt{3}$ $a_{m}$$\simeq $ 10.1 \AA
\thinspace\ $b_{m}^{\prime }$= $b_{m}$$\simeq $ 5.71 \AA \thinspace\ $%
c_{m}^{\prime }$= $2c_{m}$$\simeq $ 8.27 \AA \thinspace\ and $\protect\beta $%
$\simeq $ $145.5^{o}$ is indicated by the thick lines.}
\end{figure}

\begin{figure}[tbp]
\includegraphics[width=0.7\textwidth] {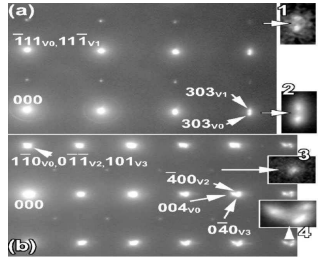}
\caption{Electron diffraction patterns in the \TEXTsymbol{<}211\TEXTsymbol{>}
(a) and \TEXTsymbol{<}110\TEXTsymbol{>} (b) zones at about 87 K. Insets 1-4
show an enlargement of the regions indicated by arrows, respectively. (a)
shows two M$_{A}$ variants, V$_{0}$ and V$_{1}$(inset 2), which have a
reflection twin relationship with ($\overline{1}$01) as their twin plane.
Both M$_{A}$ variants have superlattice spots as shown in inset 1. In (b)
the fundamental spots split into three sets of spots, which correspond to
three M$_{A}$ variants V$_{0}$, V$_{2}$ and V$_{3}$. The V$_{0}$ and V$_{2}$%
, and V$_{0}$ and V$_{3}$, both yield a reflection twin relationship with
(101) and (011) as their twin planes, respectively. Careful examination
shows that there is only one set of superlattice spots, which correspond to
the V$_{3}$ spots, indicating that only V$_{3}$ shows the minority phase.}
\end{figure}

\end{document}